\documentclass[aps,twocolumn]{revtex4}
\usepackage{amsmath,amssymb}
\usepackage{graphicx}

\newcommand{\beq}{\begin{equation}}
\newcommand{\eeq}{\end{equation}}
\newcommand{\be}{\begin{equation}}
\newcommand{\ee}{\end{equation}}

\newcommand{\md}{{\mathrm{d}}}

\begin{document}

\preprint{AEI--2004--051}

\title{Loop Quantum Gravity and the Cyclic Universe}

\author{Martin Bojowald$^1$\footnote{e-mail address: {\tt mabo@aei.mpg.de}},
Roy Maartens$^2$\footnote{e-mail address: {\tt
roy.maartens@port.ac.uk}}, Parampreet
Singh$^{3,4}$\footnote{e-mail address: {\tt param@iucaa.ernet.in}}
}

\address{~}

\address{$^1$Max-Planck-Institut f\"ur Gravitationsphysik,
Albert-Einstein-Institut, Am M\"uhlenberg 1, D-14476 Potsdam,
Germany}

\address{$^2$Institute of Cosmology and Gravitation,
University of Portsmouth, Portsmouth PO1~2EG, UK}

\address{$^3$IUCAA, Ganeshkhind, Pune 411 007, India}

\address{$^4$Astronomy Unit, School of Mathematical Sciences,
Queen Mary, University of London, London E1 4NS, UK}

\date{\today}

\begin{abstract}

Loop quantum gravity introduces strong non-perturbative
modifications to the dynamical equations in the semi-classical
regime, which are responsible for various novel effects, including
resolution of the classical singularity in a Friedman universe.
Here we investigate the modifications for the case of a cyclic
universe potential, assuming that we can apply the
four-dimensional loop quantum formalism within the effective
four-dimensional theory of the cyclic scenario. We find that loop
quantum effects can dramatically alter the near-collision dynamics
of the cyclic scenario. In the kinetic-dominated collapse era, the
scalar field is effectively frozen by loop quantum friction, so
that the branes approach collision and bounce back without actual
collision.

\end{abstract}

\maketitle

\section{Introduction}

Despite the dramatic advances in high-precision data and our
ability to tie down the cosmological parameters with growing
accuracy, there remain a number of deep unresolved puzzles in our
understanding of the universe. These include the origin of the
universe, the fundamental theory that underlies inflation -- or
that provides an alternative to inflation, and the origin and
nature of the dark energy. A bold attempt to tackle these problems
is the ekpyrotic/ cyclic scenario~\cite{ekcyc}, which invokes
ideas from M~theory to construct an alternative to the standard
inflationary paradigm.

A crucial issue for the cyclic scenario is how to process the
cosmological dynamics and perturbations through the singularity at
the instant of brane collision. One possible resolution of this
problem is that quantum gravity effects will in fact prevent a
collision, and thereby avoid a singularity in the
higher-dimensional spacetime. In the absence of a
higher-dimensional non-perturbative quantum gravity formalism, we
use a four-dimensional approach and apply it to the effective
four-dimensional description of the cyclic scenario.

The effective four-dimensional description of the cyclic scenario
involves a scalar moduli field $\varphi$ on the visible brane that
encodes the brane separation. Its effective potential $V(\varphi)$
determines the inter-brane distance, and it dominates the dynamics
on the visible brane around the time of approach to collision. We
seek to investigate possible non-perturbative (but semi-classical)
quantum corrections during this time.

Loop quantum gravity is a four-dimensional non-perturbative
candidate theory of quantization of
spacetime~\cite{rovelli_thomas}, whose successes include
prediction of a discrete spectrum for geometrical
operators~\cite{geometrical_op}, matter Hamiltonians that are free
from ultraviolet divergences~\cite{thiemann_matter} and derivation
of the Bekenstein-Hawking entropy formula~\cite{bek_hawking}.
Recently, loop quantum gravity has been applied to cosmology (for
reviews see~\cite{lqc_review,icgc}), leading to a resolution of
cosmological singularities~\cite{singularity} and a new view on
initial conditions~\cite{Initial}. In general, singularity
avoidance entails a breakdown of smooth classical spacetime
structure and the quantum geometric discretization of spacetime.
Loop quantum cosmology derives a difference equation for the wave
function whose evolution does not stop where the classical
singularity would be. The system then continues to a new branch
after which a semi-classical description may be used~\cite{Semi}.

Loop cosmology predicts that as we approach smaller scales the
classical continuous spacetime picture is replaced by quantum discrete
spacetime. However there is an intermediate region in this transition
where evolution can be described by a continuous spacetime with
non-perturbative quantum modifications. This implies that one can use
effective classical equations with a coordinate time parameter rather
than wave functions for the analysis, which simplifies the
calculations considerably. One can understand the effective classical
equations as describing the position of a wave packet moving in
coordinate time, as elaborated in more detail in \cite{Time}.  In this
regime the geometrical density has a very non-classical behavior in
the sense that it starts {\em decreasing} as the scale factor
decreases~\cite{InvScale}. For a Friedman-Robertson-Walker (FRW)
universe with a scalar field this effect changes the frictional term
to anti-frictional (or vice-versa) in the Klein-Gordon equation below
a critical scale factor~\cite{inf_martin}.

This mechanism has various interesting applications. For example,
it drives a short period of super-inflation during which the
inflaton is pushed up its potential hill~\cite{inf_martin}, thus
providing a new perspective on how initial conditions may be set
for standard inflation~\cite{inf_cmb}. The mechanism is robust to
various quantization freedoms~\cite{inf_amb} and can in principle
leave a small observational signature on the largest scales in the
cosmic microwave background anisotropies~\cite{inf_cmb}. Another
application of loop effects is the resolution of the big crunch
problem: closed collapsing FRW universes always bounce and escape
the crunch, irrespective of initial
conditions~\cite{bounce_closed}. The mechanism has also been
applied to study inflation for oscillating closed
universes~\cite{closed_inf_osc}, and to resolve chaotic behavior
and singularities in anisotropic models~\cite{bianchi}.

\begin{figure}[tbh!]
\includegraphics[width=8.5cm,height=6cm]{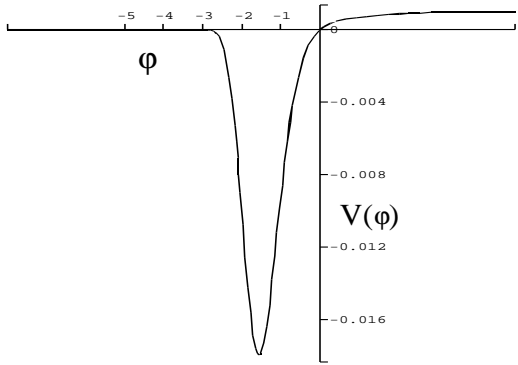}
\caption{\label{pot} Typical cyclic potential [corresponding to
parameters $V_0 = 10^{-3},  m_1 = 0.2$ and $m_2 = 20/21$ in
Eq.~(\ref{cycpot})]. }
\end{figure}

In the cyclic scenario, visible and shadow 4-dimensional branes
move in a 5-dimensional bulk, with their separation determined by
a moduli field $\varphi$, driven by a potential $V(\varphi)$ which
is slightly positive for $\varphi > 0$ and asymptotes to zero as
$\varphi \to - \infty$ (see Fig.~\ref{pot}). For $\varphi$
negative the potential is very steep and negative, turning around
and increasing as $|\varphi|$ increases. The turn-around
corresponds to close approach of the branes, at the string scale,
where quantum effects begin to dominate. In the effective theory,
$\varphi$ moves rapidly through this turn-around, and the branes
collide as $\varphi \to - \infty$. This in turn initiates a hot
radiation era on the physical brane, leading to standard
cosmological evolution as the branes separate. The cycle is
repeated as the expansion turns around on the visible brane and
collapse sets in, with the shadow brane once more approaching for
another collision.

In the 5-dimensional frame, the density and curvature on the visible
brane remain finite at collision~\cite{ekcyc}. However, the fifth
dimension itself degenerates at collision, and there is no well-posed
Cauchy problem. The scale factor vanishes at collision, but the matter
and radiation densities do not diverge at the instant of collision,
owing to the coupling of $\varphi$ with matter and radiation. This
evades the usual 4-dimensional singularity, but the problem re-appears
as a singularity in the extra dimension.  The key issue for the cyclic
scenario is to find a 5-dimensional way through the collision, so as
to be able to predict the cosmological perturbation spectra.

Our aim here is different. We are interested in the possible role
of non-perturbative quantum corrections as the branes approach,
i.e., near the steep minimum in the potential. To this end, we
investigate how 4-dimensional loop quantum effects will alter the
effective 4-dimensional dynamics of the cyclic scenario (since
there is no 5-dimensional non-perturbative formalism).

The key point is that the steep negative potential produces a
strong kinetic regime for the moduli field, and so we expect
strong loop quantum corrections to be triggered. Previous results
suggest that these corrections will prevent the singularity, i.e.,
the branes will come close but not actually collide. Essentially
we confirm this expectation. However, we also find that it is
difficult to achieve the bounce without passing from the
semi-classical regime to the high-energy fully quantum regime,
where our use of the effective 4-dimensional theory breaks down.
The problem is that the kinetic energy and the Hubble rate
typically reach Planckian scale as the branes approach, and then
we have no handle on the dynamics and are unable to make
predictions.

In the semi-classical regime where we can apply the loop
corrections, brane collision is prevented. We cannot make
predictions for the subsequent evolution since we neglect the
radiation on the brane. Our results only apply in the near-brane
regime where the moduli field is dominant.

\section{Dynamics in loop cosmology}

The main modification in the classical general relativistic
equations implied by loop quantum cosmology is that inverse powers
of the scale factor in the matter Hamiltonian are replaced by a
bounded function~\cite{InvScale,icgc}. This can be interpreted as
a curvature cut-off naturally following from the discrete
structure underlying the loop quantization. For a scalar field the
Hamiltonian is
\begin{equation}
{\cal H}=\frac{1}{2}d(a)p_{\varphi}^2+a^3 V(\varphi) \,,
\end{equation}
where $p_{\varphi}$ is the momentum canonically conjugate to
$\varphi$, and $d(a)$, which is classically $1/a^3$, encodes the
quantum corrections. In the semi-classical regime, where spacetime
may be treated as continuous, it is given by
\begin{equation}
d(a)= {D(q) \over a^3}\,,~q={a^2 \over a_*^2}\,, ~
a_*=\sqrt{{\gamma j \over 3}}\, \ell_{\rm Pl}\,.
\end{equation}
where $j$ (a half integer) is a quantization parameter, $\gamma
\approx 0.13$ is the Barbero-Immirzi parameter, $\ell_{\rm Pl}$ is
the Planck length and
\begin{eqnarray} \label{defD}
D(q) =q^{-3/2} \left\{\frac{3}{2l}\left({1 \over l+2}
\left[(q+1)^{l+2}-|q-1|^{l+2}\right]\right.\right. \nonumber\\
 - \left.\left.\frac{q}{1 + l}
\left[(q+1)^{l+1}-{\rm sgn}(q-1)
|q-1|^{l+1}\right]\right)\right\}^{3/(2-2l)}\!.
\end{eqnarray}
Here $l$ is another quantization parameter, with $0<l<1$. This
expression for $D(q)$ was derived in \cite{InvScale} for $l=3/4$ as an
approximation to the eigenvalues of inverse scale factor operators in
loop quantum cosmology, and generalized in \cite{icgc} to arbitrary
$0<l<1$. The approximation to the eigenvalues becomes better for
values of $j$ larger than the minimal one, $1/2$.

The scale below which non-perturbative modifications become
important is given by $a_*$.
Typically, one chooses $j\gg 1$, so that
$a_*\gg \ell_{\rm Pl}$. The Planck scale marks the onset of
discrete spacetime effects. For $\ell_{\rm Pl}<a \ll a_*$, we are
in the semi-classical non-perturbative regime, where the
geometrical density behaves as
 \be \label{approxD}
d(a) \sim \left[\left({3\over 1+l}\right)^{3/(2-2l)} \left({a
\over a_*}\right)^{3(2-l)/(1-l)}\right] {1\over a^3}\,.
 \ee

The Hamiltonian determines the dynamics completely. The matter
Hamiltonian leads to the Klein-Gordon equation via the Hamiltonian
equations of motion. This first equation gives
\begin{equation}
\dot{\varphi}=\{\varphi,{\cal H}\} = d(a) p_{\varphi}\,.
\end{equation}
The second Hamiltonian equation of motion for $p_{\varphi}$ can
then be recast into a second order equation for
$\varphi$~\cite{inf_martin,inf_cmb,bounce_closed},
\begin{equation}
\ddot \varphi + \left(3 H - \frac{\dot D}{D} \right) \dot \varphi
+ D \, V_{,\varphi} = 0\,. \label{kgeq}
\end{equation}
For $\ell_{\rm Pl}<a \ll a_*$, we find that $\dot D/D>3 H$, and
this leads to the classical frictional term for an expanding
universe becoming anti-frictional, or vice versa if the universe
is contracting.

The Friedman equation follows by equating the matter Hamiltonian
and the gravitational contribution $3\dot{a}^2a$,
\begin{equation}\label{fe}
H^2=\frac{8\pi G}{3}\left[\frac{\dot{\varphi}^2}{2D(a)}+
V(\varphi)\right]\,.
\end{equation}
Finally, the Raychaudhuri equation follows from a Hamiltonian
equation of motion for the gravitational Hamiltonian constraint:
\begin{equation}\label{re}
\frac{\ddot{a}}{a}= -\frac{8\pi G}{3}\left[
\frac{\dot{\varphi}^2}{D} \, \left(1-\frac{\dot{D}}{4 H D}\right)
-V(\varphi)\right]\,.
\end{equation}

The Friedman equation implies that a bounce in the scale factor,
i.e., $\dot{a}=0$ and $\ddot{a}>0$, requires a negative potential.
(In a closed model, the curvature term allows for a bounce with
positive potential~\cite{bounce_closed}.) This occurs for a
negative cosmological constant or the potential considered in
cyclic models. Vanishing Hubble parameter at the bounce implies
\begin{equation} \label{phid}
\dot{\varphi}^2={-2D(a)V(\varphi)}\,,
\end{equation}
so that at the bounce,
 \be
\frac{\ddot{a}}{a} = \frac{4\pi G}{3}\left(6- \frac{\md \ln D}{\md
\ln a}\right) V \,.
 \ee
Thus classically, i.e., for $D= 1$, a bounce for a negative
$V(\varphi)$ is not allowed. With the modified $D(a)$, however, $
\md\ln D/\md \ln a>6$ will hold for sufficiently small $a$, so
that $\ddot{a}>0$ is possible. Thus, the universe has to collapse
sufficiently deep into the modified regime before it can bounce
back. In this regime, $D(a)$ is decreasing with shrinking scale
factor, so that $\dot{\varphi}^2$ around the bounce, given by
Eq.~(\ref{phid}), is very small. Thus the scalar field almost
freezes, which can also be seen as a consequence of the loop
quantum friction effect in the Klein-Gordon equation for a
contracting universe. In general relativity, the corresponding
term is strongly anti-frictional during collapse, so that
$\dot\varphi^2$ increases and no turn-around is possible.

After the bounce, $\varphi$ unfreezes and continues its motion
with $|\dot{\varphi}|$ becoming larger. Another consequence of the
Friedman equation with a negative scalar potential, is that
$\dot{\varphi}$ cannot become exactly zero, so that $\varphi$ will
not turn around and just slow down during the bounce. The bounce
in scale factor is not a bounce in $\varphi$. Some time after the
bounce, the universe may recollapse, which also is possible only
for a negative potential. In contrast to the bounce, however, this
is a purely classical effect, since the volume has become large.

\section{Loop quantum effects for a cyclic potential}

We start with a simple illustration for the case of constant
negative potential (or a negative cosmological constant), which
will be followed by a typical potential used in the cyclic
scenario.

\subsection{Negative cosmological constant}

For a free massless scalar field in a constant negative potential
$V=\Lambda/8\pi G\equiv \lambda$, the Klein-Gordon
equation~(\ref{kgeq}) can be integrated once to obtain
$\dot{\varphi}=C D(a)/a^3$, where $C$ is a constant. Inserting
this into the Friedman equation~(\ref{fe}) gives
\begin{equation}\label{a}
\dot{a}^2-\frac{8\pi G}{3}\left[C^2 \frac{D(a)}{2 a^4} + \lambda
a^2\right]=0\,.
\end{equation}
The evolution of the scale factor is determined by the effective
potential
 \be \label{veff}
V_{\rm eff} = -\frac{C^2}{2}\frac{D(a)}{a^4}-\lambda a^2\,.
 \ee

This potential already demonstrates the difference between the
classical and the loop quantum cases. Classically, the first term
dominates at small $a$, so that $V_{\rm eff}\to-\infty$. Thus
$a\to0$ and a singularity follows. With the quantum modified
$D(a)$, however, the potential approaches zero at $a=0$, and there
is a barrier of positive potential at small $a$. At intermediate
$a$, extending from the modified region to large $a$, there is a
classically allowed region which describes the periodic motion of
the scalar. At large $a$ there is a turning point in both the
classical and the effective case.

Since the maximal scale factor $a_{\rm max}$ lies in the classical
regime, we have $D(a_{\rm max})\approx1$ at $V_{\rm eff}(a_{\rm
max})=0$, so that $a_{\rm max}\approx (-C^2/2\lambda)^{1/6}$. The
minimum scale factor $a_{\rm min}$ lies in the modified regime.
For $(\ell_{\rm Pl}<)a_{\rm min}\ll a_*$, we can use
Eq.~(\ref{approxD}) to obtain $D(a_{\rm min})\approx B
a^{3(2-l)/(1-l)}$, where $B\equiv [3/(1+l)]^{3/(2-2l)}$. By
Eq.~(\ref{veff}), $a_{\rm min}\approx
(-2\lambda/BC^2)^{(1-l)/3l}$. Thus, the cosmological constant must
be small enough, $|\lambda|<C^2B^{(2-2l)/(2-l)}/2$ for a
consistent solution, i.e., $a_{\rm min}<a_{\rm max}$.

The equations of motion can be solved in this case up to
integrations. From Eq.~(\ref{a}) we obtain $t(a)=\int\md a
[-V_{\rm eff}(a)]^{-1/2}$, which upon inversion leads to $a(t)$ so
that $\dot{\varphi}=C D(a)/a^3$ can be integrated. The resulting
motion for $a$ is periodic, while $\varphi$ changes monotonically
(with periodic $\dot{\varphi}$). The period is given by
 \begin{eqnarray}
T &=& 2\int_{a_{\rm min}}^{a_{\rm max}}\frac{\md a}{\dot{a}}
\nonumber\\ &=& \sqrt{3}\int_{a_{\rm min}}^{a_{\rm max}} \frac{\md
a}{\sqrt{\pi G(C^2a^{-4}D(a)-2|\lambda|a^2)}}\,.
 \end{eqnarray}

\subsection{Cyclic potential}

When the potential is negative but non-constant, the scale factor
still behaves cyclically, but non-periodically. The size of the
scale factor at recollapse changes between cycles in a way
depending on the potential. This case is in particular relevant
for cyclic models which consider the potential (see
Fig.~\ref{pot})
 \be \label{cycpot}
V(\varphi) = V_0 \, (1 - e^{- \varphi/m_1})
\exp(-e^{-\varphi/m_2})\,,
 \ee
where $m_i$ are two energy scales, and we are using Planck units.
This models an inter-brane attractive potential for the moduli
field $\varphi$, which describes the brane separation via
$e^{\varphi}$, so that the branes collide for $\varphi\to-\infty$.
At the same time, the scale factor approaches zero following the
classical equations of motion. However, the matter and radiation
densities on the brane do not diverge.

This scenario critically relies on unknown and non-classical
dynamics which occurs when branes collide and then re-emerge to
move apart. Non-perturbative quantum corrections are needed to
resolve this issue, but no suitable 5-dimensional formalism is
available. Instead, one can try to apply the 4-dimensional loop
quantum formalism as a non-perturbative correction to the
effective 4-dimensional theory of the cyclic scenario.

\begin{figure}[tbh!]
\includegraphics[width=8.5cm]{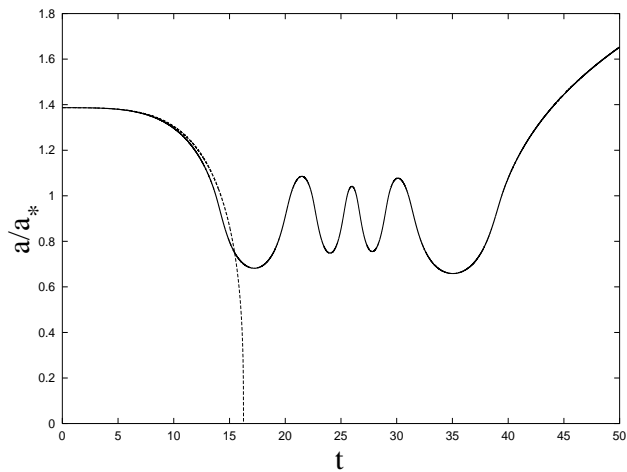}
\caption[]{\label{scale_fig} The evolution of the scale factor
$a(t)$ from a small initial value, for the cyclic potential. The
dashed curve shows the standard case, while the solid curve
incorporates loop quantum corrections, which stop the branes from
colliding. The loop parameters are $j = 120$ and $l =3/4$ and the
cyclic parameters are $V_0 = 10^{-3}$, $m_1= 0.2$, $m_2 =20/21$.}
\end{figure}

\begin{figure}[tbh!]
\includegraphics[width=8.5cm]{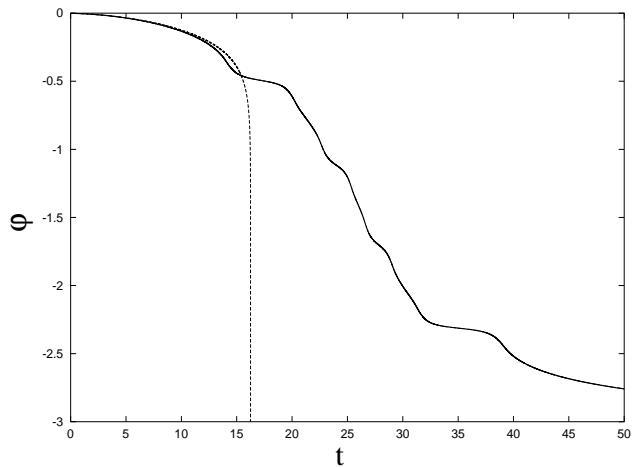}
\caption[]{\label{phi_fig} Evolution of the moduli field
$\varphi(t)$ for the case described in Fig.~\ref{scale_fig}.}
\end{figure}

\begin{figure}[tbh!]

\includegraphics[width=8.5cm]{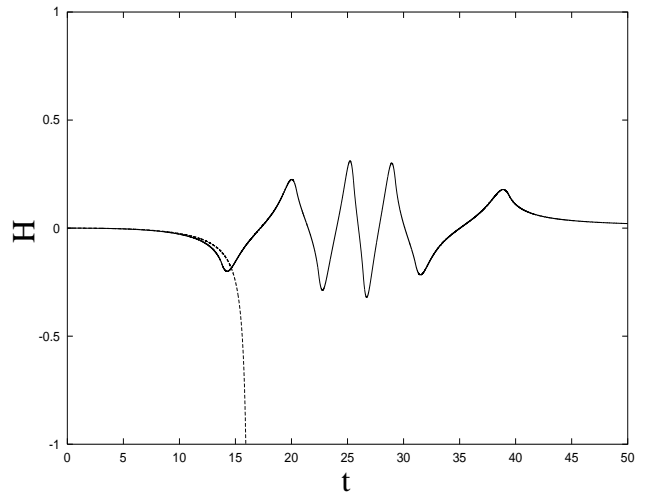}
\caption[]{\label{hub_fig} Evolution of the Hubble rate $H(t)$ for
the case described in Fig.~\ref{scale_fig}.}
\end{figure}

We now have a detailed prescription for quantum gravity effects in
the semi-classical regime which modify the scalar field dynamics
describing this situation. Before $\varphi$ reaches $-\infty$, the
scale factor enters the regime where the frictional term in the
modified Klein-Gordon equation slows down the scalar $\varphi$
while $a$ bounces back. Subsequently, $\varphi$ does not turn
around but continues to decrease. The comparative evolution of the
scale factor, moduli field and Hubble rate, with and without loop
quantum effects, is illustrated in
Figs.~\ref{scale_fig}--\ref{hub_fig}. The classical cyclic
universe has $a\to 0$ in a finite time, whereas for the same set
of parameters but with loop quantum corrections, the universe
bounces back before a singularity occurs. In the standard cyclic
scenario, as $\varphi$ runs down the steep negative potential, the
Hubble parameter exceeds the Planck energy in the 4-dimensional
frame, but this can be avoided via loop quantum corrections if the
initial scale factor at the onset of collapse is small
enough~\cite{rn} (see the solid curve in Fig.~\ref{hub_fig}
compared to the classical dashed curve).

Since the potential is not constant, the behavior of expansion and
recollapse of the scale factor will not be periodic but will still
be cyclic. The constant potential indicates that the smaller the
magnitude of the potential, the larger the expansion between a
bounce and the following recollapse, which is also suggested by
the figures. However, we are unable to draw definite conclusions
about the behaviour after the first avoidance of collision, since
we have neglected the matter and radiation on the visible brane,
so that we cannot predict the late-time expansion and subsequent
recollapse. From the quantum modification in the near-brane regime
where $\varphi$ dominates, it follows that the bounce is
non-singular and the singularity is removed. Note that there is a
series of bounces at finite values of $\varphi$, and due to the
slowing down of $\varphi$ during bounces, the moduli field does
not reach $-\infty$ as in the standard case (see the solid curve
in Fig.~\ref{phi_fig}). In this sense $\varphi\to-\infty$ is not a
singularity but rather a boundary at infinity.

\section{Conclusions}

The problem of unknown dynamics near brane collisions has been a
constraint on developments for the cyclic scenario. It is expected
that quantum gravity will resolve this issue, but no fully
non-perturbative higher-dimensional formalism has been available for
tackling this problem. In the absence of such a formalism, we have
applied loop quantum gravity to find non-perturbative corrections (in
the semi-classical regime, $\ell_{\rm Pl}<a\ll a_*$) to the effective
4-dimensional theory of the cyclic scenario, using a typical
potential. These non-perturbative corrections lead to very different
dynamics compared to the standard classical effective dynamics.

We have considered the regime where branes are close to each other and
the moduli field is dominant over matter and radiation, which we have
neglected. As the branes approach each other and the moduli field runs
down the steep negative potential, the kinetic term becomes dominant
and the loop quantum effects freeze the field. Classically, in the
4-dimensional frame the cyclic universe would have gone into a big
crunch with the Hubble parameter rapidly becoming much bigger than
Planck energy (in the 5-dimensional frame this corresponds to brane
collision with no singularity on the branes~\cite{ekcyc}). However,
the loop quantum gravity effects avoid the big crunch. The Hubble
parameter can be kept below the Planck energy by choosing the initial
scale factor small enough. If the scale factor is much larger than
$a_*$ at the instant when collapse starts, then the Hubble parameter
will violate the Planck bound, and we cannot apply the semi-classical
equations used here, since spacetime becomes discretized. This is an
unavoidable restriction for the potential considered, and shows that
fully quantum gravity dynamics are needed to study the general
problem. However, our limited results suggest that avoidance of brane
collision may be a general feature. Moreover, since in the cases
studied here the bounce value of the scale factor lies well above the
Planck regime, we can trust the effective semi-classical equations
used throughout, and do not have to refer to a more complicated
analysis in terms of wave functions.

It would be interesting to see whether further input from string/
M~theory leads to similar conclusions. This would introduce new
ingredients via the intrinsic properties of branes as quantum
objects. In order to make a more meaningful comparison however, it
would be necessary to go beyond the 4-dimensional approach used
here, and derive a 5-dimensional loop quantum gravity scheme for
the full bulk spacetime. The spatial inhomogeneity of the bulk
represents a considerable challenge to this project. But the
importance of the project goes beyond the cyclic scenario -- a
higher-dimensional loop quantum gravity formalism may provide the
basis for useful interaction between this approach and the string
theory approach to quantum gravity.

~\\{\bf Acknowledgements:} We thank Jim Lidsey, David Mulryne, Nelson
Nunes and Reza Tavakol for helpful discussions. RM's work is supported
by PPARC. PS thanks Max-Planck-Institut f\"ur Gravitationsphysik,
Portsmouth University and Queen Mary, University of London for
supporting visits and for warm hospitality during various stages of
this work.

\end{document}